\documentclass[aps,prl,amsmath,amssymb,superscriptaddress,showpacs,twocolumn]{revtex4-1}

\usepackage{graphicx} \usepackage{color} \usepackage{verbatim} 

\def\eV{\mathrm{eV}}

\def\PB{PuB$_6$\,}
\def\SB{SmB$_6$\,}

\begin{document}

\title{Plutonium hexaboride is a correlated topological insulator
}

\author{Xiaoyu Deng}
\affiliation{Department of Physics and Astronomy, Rutgers University, Piscataway, NJ 08854}
\author{Kristjan Haule}
\affiliation{Department of Physics and Astronomy, Rutgers University, Piscataway, NJ 08854}
\author{Gabriel Kotliar}
\affiliation{Department of Physics and Astronomy, Rutgers University, Piscataway, NJ 08854}

\date{\today}

\begin{abstract}
  We predict that plutonium hexaboride (\PB) is a strongly correlated
  topological insulator, with Pu in an intermediate valence state of
  Pu$^{2.7+}$.  Within the combination of dynamical mean field theory
  and density functional theory, we show that \PB is an insulator in
  the bulk, with non-trivial $Z_2$ topological invariants. Its
  metallic surface states have large Fermi pocket at $\bar{X}$ point
  and the Dirac cones inside the bulk derived electronic states
  causing a large surface thermal conductivity. \PB has also a very
  high melting temperature therefore it has ideal solid state
  properties for a nuclear fuel material.
\end{abstract}

\maketitle

Analogies between 4$f$ and 5$f$ materials have proved to be a fruitful
source of insights and have led to the discovery of important new
classes of materials with remarkable properties. The 5$f$ electrons in
the actinides have substantially larger relativistic effects and the
increased ionic radius of the 5$f$ electrons enhances effects
associated to the delocalization-localization transition. For example,
the alpha to delta transition in plutonium, a 5$f$ analog of the
famous volume collapse transition in Cerium, is the largest volume
change in an elemental solid, with a volume change of the order of
30$\%$ ~\cite{Pu-VolumeCollapse}.  A second noteworthy example is
provided by the enhancement of the superconducting transition in 115
compounds.  PuCoGa$_5$, a 5$f$ analog of the Ce 115 heavy fermion
superconductors, has the highest superconducting transition
temperature of $18\,$K~\cite{sarrao_plutonium-based_2002}, among all
the heavy fermion superconductors \cite{SCinHF-review-Sarrao}.

Motivated by the discovery of topological insulating state
\cite{TI-RMP-Qi,TI-RMP-Hasan} in strongly correlated samarium
hexaboride (\SB), which has been the subject of recent interest both
theoretically\cite{SB-Takimoto,TopologicalKI-Dzero,TopologicalKI-PRB-Dzero,SB_Lu,cubicTKI-arxiv-Alexandrov}
and
experimentally\cite{Miyazaki-Momentum-2012,SB-trueTI-Wolgast,zhang_hybridization_2013,
  kim_robust_2012,xu_surface_2013,SB-ARPES-Neupane,li_quantum_2013,SB-ARPES-Jiang,SB-dopedtransport-Kim,thomas_weak_2013,frantzeskakis_kondo_2013,yee_imaging_2013},
we search for analog material among 5$f$ compounds since going from
4$f$'s to 5$f$'s increases the $f$-$f$ overlap and the resulting
energy scales.  We identify plutonium hexaboride (\PB) as a strongly
correlated topological insulator at low temperatures and we
investigate its physical properties using first principles methods.
The theoretical information, combined with its experimentally known
exceptionally high melting point~\cite{B-Pu-review}, suggest that \PB
has interesting solid state properties, desired for nuclear fuels.

The topological nature of an insulator, and thus the existence of
topologically protected surface states, is described by topological
invariants\cite{TI-Z2-Kane,TI-inversionsymmetry-Fu,TI-Z2-Moore,TI-fieldtheory-Qi,TI-Z2interacting-Wang}.
Most TIs found so far are weakly correlated band insulators
\cite{TI-inversionsymmetry-Fu, HgTe-science-Bernevig,Bi2Se3-NP-Zhang}
where topological Z$_2$ invariants can be found by considering all
occupied bands
~\cite{TI-Z2-Kane,TI-inversionsymmetry-Fu,3DTI-Fu,TI-Z2-Moore,Bi2Se3-NP-Zhang,
  TI-Firstprinciple-Zhang}.  For interacting systems the proper
topological invariants are defined in terms of single-particle Green's
functions according to topological field
theory\cite{TI-Z2interacting-Wang, TI-Z2interactingPRB-Wang,
  TI-monopole-Qi}. Currently there are only a few applications to
realistic materials where correlation effects are taken into account
with methods beyond density functional theory such as local density approximation (LDA)+U
and LDA+Gutzwiller method\cite{TI-Actinides-Zhang,SB_Lu}.

The LDA plus dynamical mean field theory (DMFT) method  has
proven a power framework to study the electronic structures of
correlated systems\cite{LDADMFT-RMP-Gabi,LDADMFT-review-Held}. In this
letter, we apply this method to \PB and show that it is a strongly
correlated topological insulator by computing the topological
invariants within the DMFT framework.

The LDA+DMFT calculation is performed in the charge self-consistent
implementation described in Ref.~{\cite{LDADMFT_Haule}}, which is
based on WIEN2k package{\cite{wien2k}}. We use projectors within the
large (10$\,$eV) window, and with the screening already included in
this all-electron method, the Coulomb interaction is $U=4.5\eV$, as
previously determined in Ref.~\onlinecite{Pu-valence-Shim}.  The other
Slater integrals ($F_2$, $F_4$, and $F_6$) are screened even less, and
are computed from atomic physics program \cite{cowan} and re-scaled to
80\% of their atomic value.  To solve the impurity problem, we use
continuous-time quantum Monte-Carlo method with hybridization
expansion~\cite{CTQMC_Haule,CTQMC_Werner}. The Brillouin zone
integration is performed with a regular 10x10x10 mesh. The muffin-tin
radius is 2.50 and 1.61 Bohr radius for Pu and B, respectively and
$R_{MR}K_{max}$ is chosen to be 8.5 to ensure convergence.

\begin{figure}
\includegraphics[trim=20mm 45mm 10mm 40mm, clip=true, width=0.95\columnwidth]{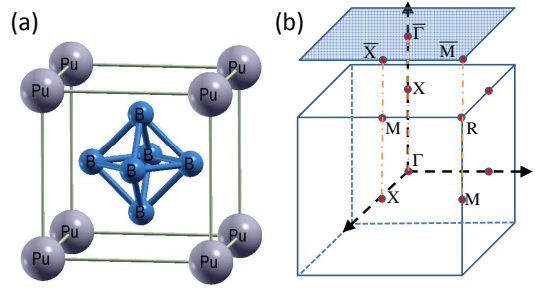}
\caption{(Color online). (a) The crystal structure of \PB with $Pm\bar{3}m$ space
  group and (b) the corresponding first Brillouin Zone for bulk and
  surface (001).}
\label{structure} 
\end{figure}
\begin{figure}
\includegraphics[width=0.95\columnwidth]{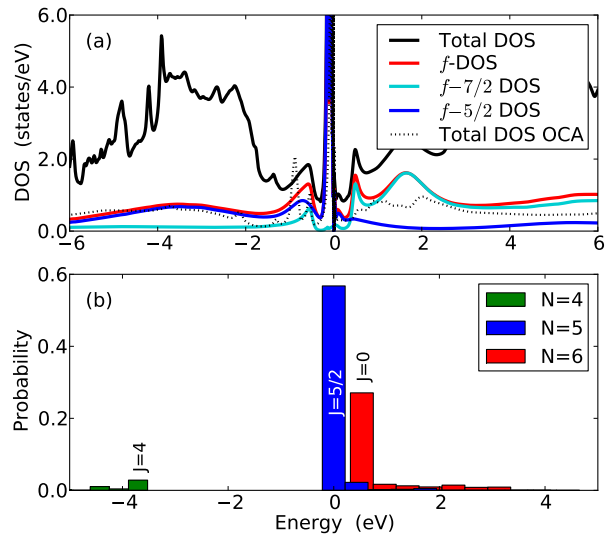}
\caption{(Color online). The correlation signatures of \PB computed
  with LDA+DMFT. (a) Density of states (momentum-integrated spectra)
  with all characters (black), Pu-$f$ character (red),
  Pu-$f$-$5/2$-only character (blue) and Pu-$f$-$7/2$-character
  (turquoise). The total density of states obtained with one-crossing
  approximation is shown with dots. (b) Valence histogram of \PB
  computed by projecting LDA+DMFT solution onto Pu atomic
  eigenstates. The height of the bar is the probability (percentage of
  time) of Pu staying on each atomic state. The $x$-axis is the
  relative energy of each state to the lowest-energy atomic
  eigenstates. The atomic eigenstates are labeled by total electron
  number $N$ and total angular momentum $J$.}
\label{DOS_VH} 
\end{figure}

\PB is among the several binary compounds formed in plutonium-boron
systems, of which most properties are still
unknown~\cite{B-Pu-review}. It crystallizes in the same CsCl-type
structure as \SB with Pu at the corner and B$_{6}$ cluster at the
center of the cubic unit cell, as shown in
Fig.~\ref{structure}(a). The total density of states of \PB and its
projection to plutonium atom are shown in in Fig.~\ref{DOS_VH}(a). The
strong correlations effect is clearly visible in the distribution of
the Pu-$f$ spectra: a narrow quasiparticle peak at the Fermi level,
and characteristic two peak Pu satellites, which are the quasiparticle
multiplets~\cite{PuTe-Yee-2010} of plutonium. Due to maximum entropy
method used here to analytically continue Monte Carlo data to the real
axis, the satellite in the $5/2$ density of states is somewhat
overbroadened, precluding the clear separation between the $5/2$ and
$7/2$ satellite, usually seen in Pu and its
compounds~\cite{PuSe-Gouder-2000,PuTe-Durakiewicz-2004}. To affirm the
presence of these quasiparticle multiplets, we therefore also show the
density of states obtained by the one-crossing approximation
method~\cite{LDADMFT-RMP-Gabi}, which is directly implemented on the
real axis, and is more precise at higher frequency.  The main
quasiparticle peak is mainly of Pu-$f$-$5/2$ characters, and contains
only a small fraction of the total spectral weight, with quasiparticle
renormalization amplitude $Z=(1-\partial
\Sigma(\omega)/\partial\omega)^{-1}\simeq 0.2$.  We notice a reduction
of the density of states at the Fermi level indicating the formation
of a small gap.

In Fig.~\ref{DOS_VH}(b), we show the valence histogram, which
illustrates the probability of an electron on Pu to be found in any of
the atomic states of Pu atom.  The probability is obtained by
projecting the LDA+DMFT ground state of the solid onto the Pu atomic
states, which are labeled by their quantum number $\vert N,J\rangle$
where $N$ is the total electron number and $J$ is the total angular
momentum, while all the other quantum numbers are traced over. Clearly
Pu $f$-electron is restricted mainly to $\vert 5,5/2\rangle$ and
$\vert 6,0\rangle$, which highlights the strongly correlated nature of
this compound.  The $f$-electron fluctuates fast between $f^5$ and
$f^6$ atomic configuration, which results in a mixed-valence nature of
\PB, with $n_f=5.3$. This is close, but slightly more mixed valent
that the elemental Pu~\cite{Pu-valence-Shim}.  The mix-valent nature
suggests a strong screening of the magnetic moments, therefore we
expect that \PB has nonmagnetic ground state, in agreement with the
theoretical calculation.

\begin{figure}
    \includegraphics[width=0.95\columnwidth]{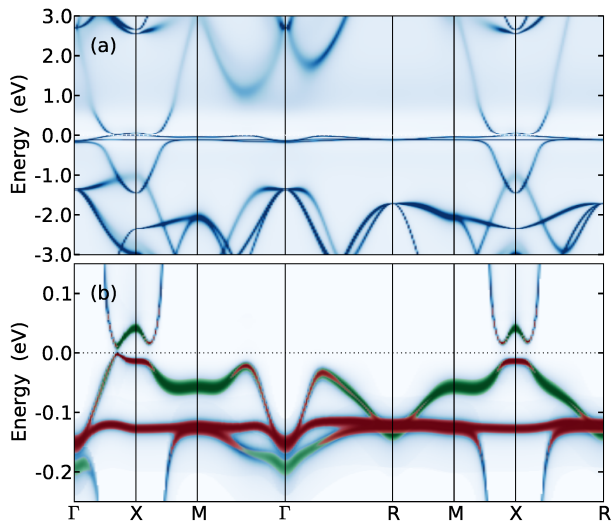}
\caption{ (Color online). The momentum-resolved spectra of \PB computed within
  LDA+DMFT method in a broad energy range (a), and a corresponding
  zoom-in around Fermi level (b) where we also depict different
  character of $f$-orbitals. Spectra with Pu-$f$-$\Gamma^-_8$
  character, Pu-$f$-$\Gamma^-_7$ character are indicated by red, green
  respectively. The corresponding basis functions are
  $\Gamma_8^{(1)}=\sqrt{\frac{5}{6}}\vert \pm \frac{5}{2}\rangle
  +\sqrt{\frac{1}{6}} \vert \mp \frac{3}{2}\rangle$,
  $\Gamma_8^{(2)}=\vert \pm \frac{1}{2}\rangle$ for the $\Gamma^-_8$
  quartet and $\Gamma_7=\sqrt{\frac{1}{6}}\vert \pm \frac{5}{2}\rangle
  -\sqrt{\frac{5}{6}} \vert \mp \frac{3}{2}\rangle$ for the
  $\Gamma^-_7$ doublet.  }
\label{ldadmftbands} 
\end{figure}

The momentum-resolved spectra $A(k,\omega)$ is shown in
Fig.~\ref{ldadmftbands}(a). Nearly flat quasiparticle bands are
located at the Fermi level with an overall bandwidth of about $0.15
\eV$, and lighter bands further away from the Fermi
level. Consistent with the density of states, the quasiparticle bands
have mainly $f$-$5/2$ characters and the spectra of $f$-$7/2$
character is pushed away from the Fermi level.  A broad band, which is
mainly of Pu-$d$ character, crosses all $f$-derived states in the
vicinity of $X(\pi,0,0)$ point, resulting in a band inversion between
$d$-orbitals and $f$-orbitals. This band inversion implies charge
transfer from Pu $f$-orbitals to Pu $d$-orbitals and is consistent
with the mixed-valent nature revealed in the valence histogram. By
examining the detailed structure of the quasiparticle bands near the
Fermi level (Fig.~\ref{ldadmftbands}{b}), we find clearly that a small
gap opens in the vicinity of the $X$-point, making \PB a narrow gap
semiconductor.  In cubic lattice environment, the $f$-$5/2$ orbitals
are split into two levels: a quartet with $\Gamma^-_8$ symmetry and a
doublet with $\Gamma^-_7$ symmetry. The orbital character is depicted
in Fig.~\ref{ldadmftbands}{b}. We see that both $\Gamma^-_8$ and
$\Gamma^-_7$ states substantially contribute to the bands near Fermi
level, and both have to be taken into account in modeling this
compound.

\begin{figure}
\includegraphics[width=0.95\columnwidth]{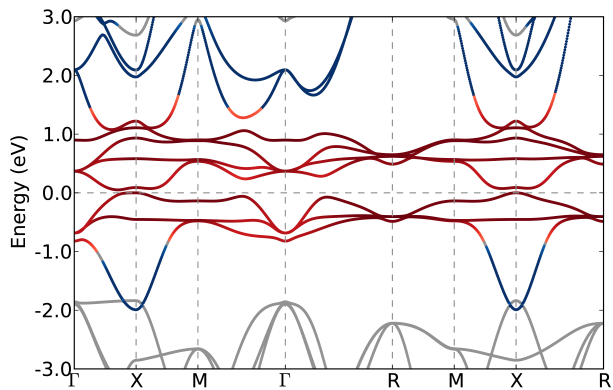}
\caption{(Color online). The band structure of \PB computed by
  LDA. The relative weight of Pu-$f$-character and $d$-bands is
  labeled by colors: red is mainly $f$-character and and blue is
  mainly $d$-character. Other characters are indicated by gray.}
\label{fig:ldabands} 
\end{figure}

We also check the LDA band structure of \PB for comparison, as shown
in Fig.~\ref{fig:ldabands}. We note that LDA also predict the
insulating state for \PB.  The main differences between
the two theoretical methods is that the $f$ derived states are much
broader within LDA (about $0.5\,$eV compared to $0.15\,$eV in DMFT),
and the $f$-7/2 states are centered just above the Fermi level in LDA,
in contrast to LDA+DMFT.  Despite these differences, the low energy
gap structure of the two methods is quite similar near the Fermi
level, with only slightly larger gap in LDA, and a direct gap in
LDA+DMFT and slightly indirect in LDA. Notice also that the high
frequency spectra of mostly light bands ( below -1eV and above 2eV)
are very similar in both methods.

\begin{figure*}
\includegraphics[scale=0.28]{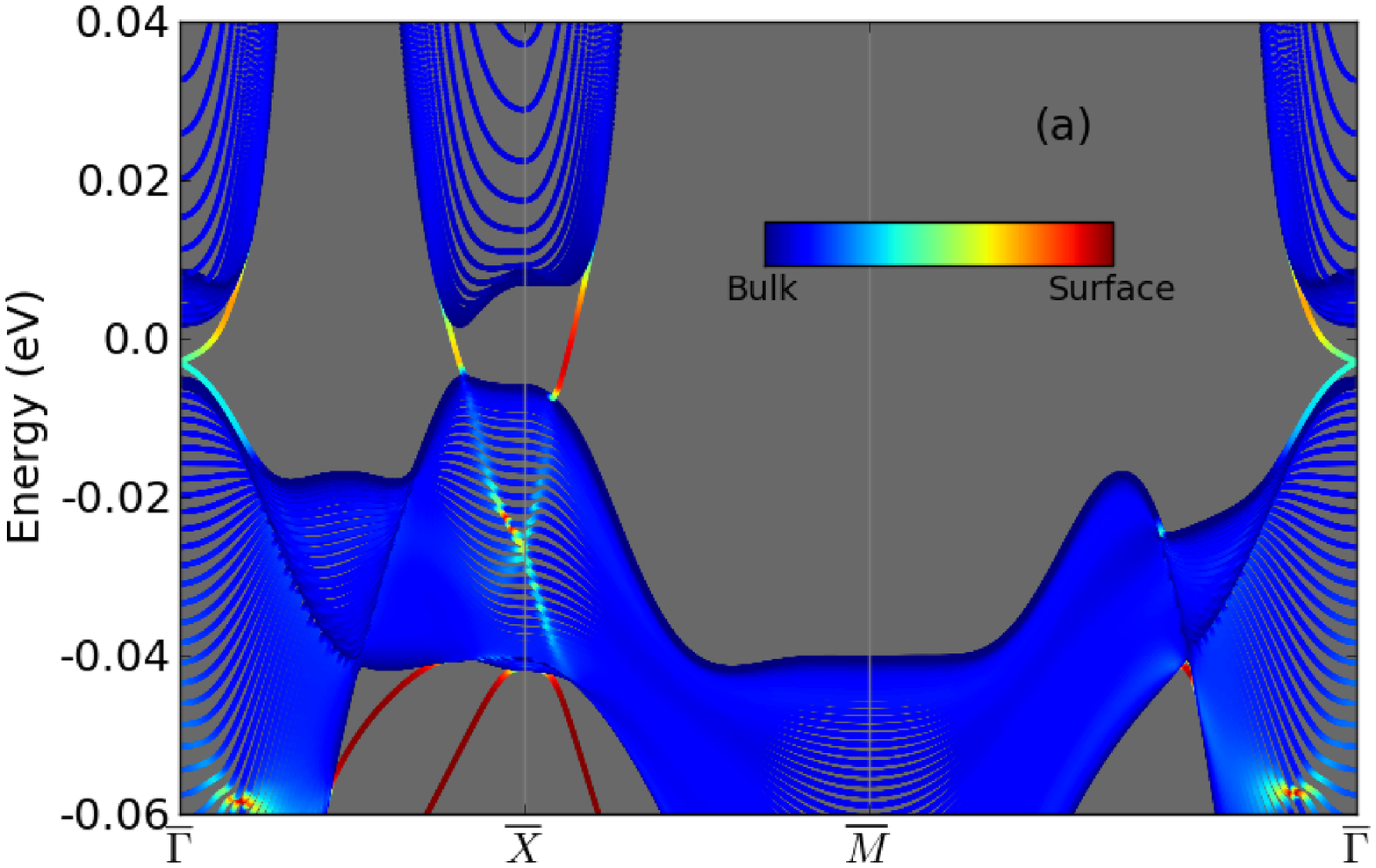}
\includegraphics[scale=0.28]{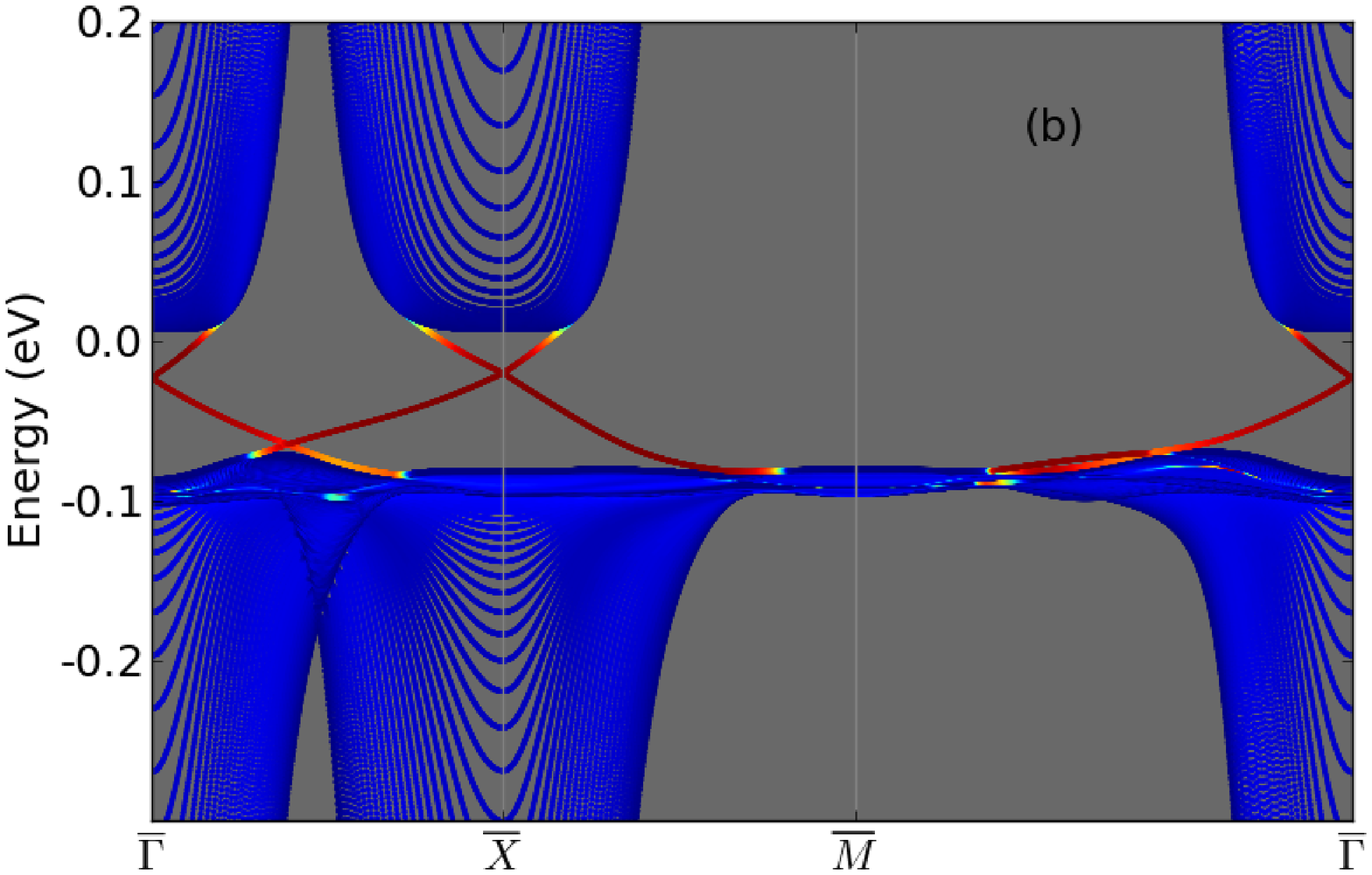}
\includegraphics[scale=0.28]{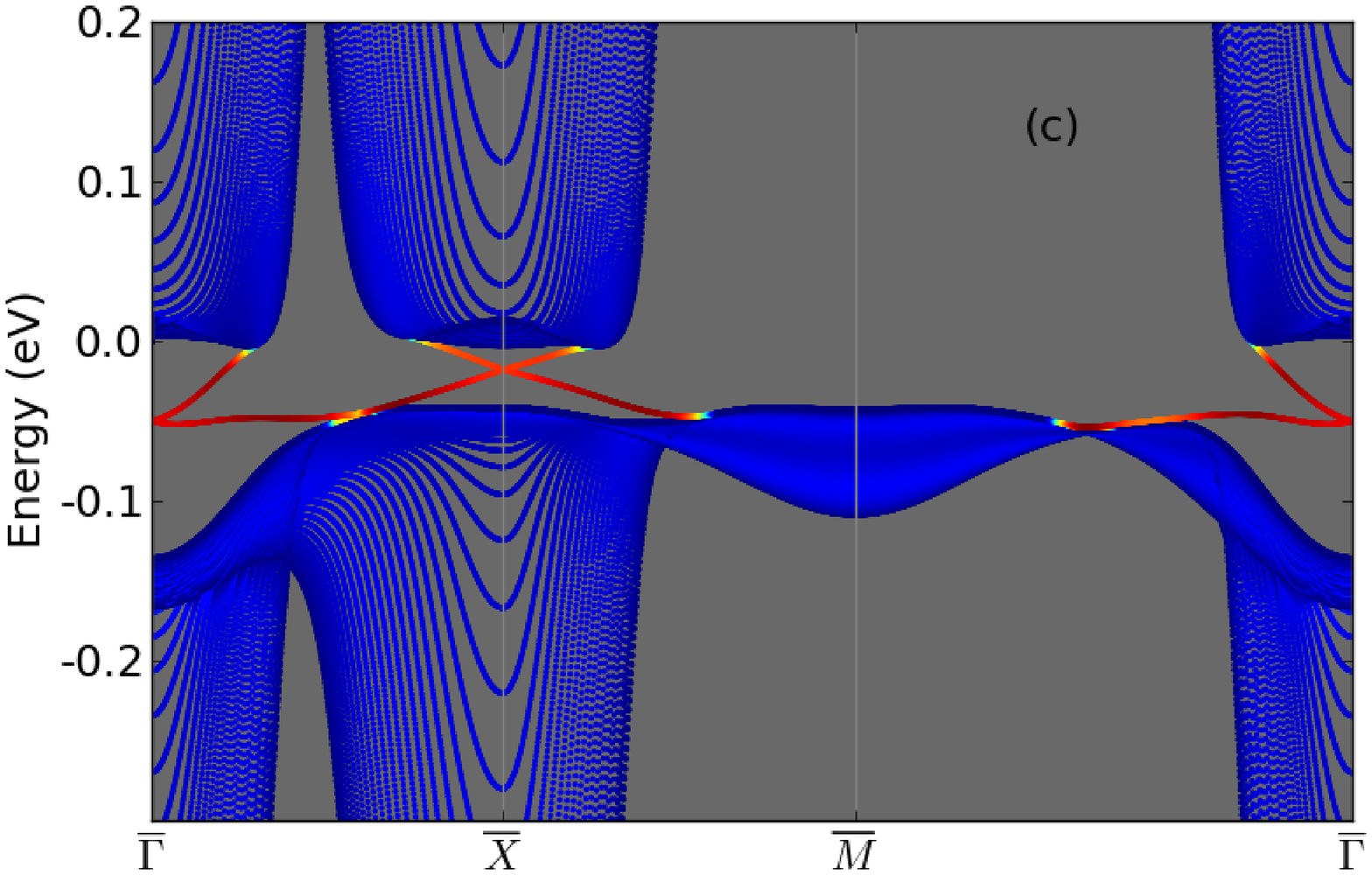}
\caption{(Color online). Surface states of \PB on (001) surface. The band structure
  is calculated with a tight-binding model for a 60-layer slab
  constructed from the effective topological Hamiltonian including
  Pu-$d$-$e_g$ orbitals and Pu-$f$ orbitals (a). The weight of surface
  state (the probability of each state in the first two layers on
  surface) are indicated by color, red means more weight on the
  surface, blue means more weight in the bulk. The same calculation is
  done with effective model Hamiltonian in which only $f$-$\Gamma^-_8$
  orbitals or only $f$-$\Gamma^-_7$ orbitals are included in additional
  to $d$-$e_g$ orbitals, and the corresponding surface states are shown in
  (b), (c) respectively.}
\label{fig:edgestates} 
\end{figure*}

After establishing the fact that \PB is a correlated mixed valent
insulator, we turn to its topological nature. To compute the
topological invariant of an interaction system, we follow
Ref.~\onlinecite{TI-Z2interacting-Wang}.  Here the self-energy has no
singularity in the vicinity of the Fermi level, hence only the Green's
function at zero frequency $G^{-1}(k,0)=\mu -H(k)-\Sigma(k,0)$ is
needed to determine the topology of the quasiparticle states.
Following
Ref.~\cite{TI-TopoHamiltonian-Wang,TI-Z2interactingPRB-Wang}, we hence
compute the topological invariant for a non-interacting Hamiltonian
defined by $H_{t}(k)=H(k)+\Sigma(k,0)-\mu$. Owing to the inversion
symmetry of \PB, we only need to compute the parity of all occupied
states~\cite{TI-Z2-Kane,TI-inversionsymmetry-Fu,3DTI-Fu}.  This method
can be applied to LDA+DMFT calculation, which is the same as the
approaches used in slave-boson , LDA+Gutzwiller and LDA+U studies
\cite{TI-Actinides-Zhang,TopologicalKI-Dzero,TopologicalKI-PRB-Dzero,SB_Lu,cubicTKI-arxiv-Alexandrov}.
In \PB there are four independent TRIMs: $\Gamma(0,0,0)$,
$X(0,0,\pi)$, $M(0,\pi,\pi)$, $R(\pi,\pi,\pi)$. Following above
technique the counted parity products of TRIMs are presented in Table
\ref{tab_parity}. This is consistent with the naive observation that
there is band inversion of $d$-orbitals and $f$-orbital at $X$-point
which contributes a parity change of -1, while for the other TRIM
points, there is no band inversion. The $Z_2$ topological index is
$(1:111)$. We conclude that \PB is a strong topological insulator in
the prediction of LDA+DMFT method and it belongs to the same
topological class as \SB \cite{SB_Lu}.

\begin{table}[h]
\begin{center}
\begin{ruledtabular}
\begin{tabular}{c c c c}
  $\Gamma$ & 3$X$ & 3$M$ & $R$ \\ \hline +1 & -1 & +1 & +1 \\
\end{tabular}
\end{ruledtabular}
\end{center}
\caption{The product of parities eigenvalues computed from occupied states of the topological Hamiltonians at
  eight TRIMs in the Brillouin zone.}
\label{tab_parity}
\end{table}

Topologically protected surface states emerge as a consequence of the
nontrivial topological nature of the bulk system. To study the surface
states, we first construct the maximally localized Wannier functions
\cite{Wannier90,W2W-Kunes,MLWF-RMP}.  At low energy, the computed
self-energy can be approximated by a Fermi-liquid-like form
$\Sigma(k,\omega)\approx\Sigma_k(0)+(1-1/Z_k)\omega$ (the imaginary
part is ignored), which leads to an effective topological
Hamiltonian given by
\[H_t^{eff}=\sqrt{Z_k}(H(k)+\Sigma_k(0)-\mu_f)\sqrt{Z_k},\] where
$\sqrt{Z}$ and $\Sigma$ are diagonal matrices in orbital space,
vanishing on the Pu-$d$ orbital.
This is a quasiparticle Hamiltonian, which accurately reproduce the
quasiparticle bands at low frequency. We then expect that the surface
states of quasiparticles can be captured by this effective topological
Hamiltonian. We solve the tight-binding model on a slab constructed
with this Hamiltonian. As shown in Fig.~{\ref{fig:edgestates}}(a), we
find that gapless edge states show up in the bulk gap around
$\bar{\Gamma}$ and $\bar{X}$, consistent with the nontrivial
topological invariants.  In our results the Dirac point at $\bar{X}$
is deep inside the bulk states, and consequently the Fermi surface at
$\bar{X}$ point is large.

Recent studies postulate minimum model Hamiltonians for such $d$-$f$
hybridization system were based on the assumption that crystal field
splitting are large enough that only one crystal level of $f$-orbitals
is relevant
\cite{SB-Takimoto,TI-Actinides-Zhang,cubicTKI-arxiv-Alexandrov}.  To
understand the effects of such artificial enhancement of the crystal
field splitting on the surface states, we construct two alternative
low energy models, keeping either $\Gamma^-_8$ or $\Gamma^-_7$ states
only.  The corresponding surface states are shown in
Fig.~\ref{fig:edgestates}(b) and (c). In both models we obtain much
larger gaps. There are gapless edge states in agreement with previous
model calculations, although the detailed dispersion of edge states
are somewhat different due to details in the hopping parameters. Three
Dirac cones show up in the gap at $\bar{X}$ and $\bar{\Gamma}$ in
consistent with previous reports
\cite{cubicTKI-arxiv-Alexandrov,SB_Lu}. Therefore we conclude that
while the existence of gapless edge states is protected by the
topology, the details of the edge states depends sensitively on the
chosen crystal field splittings in model Hamiltonians.

Recently \SB, which within LDA+DMFT method is similar to \PB, has been
studied intensively to understand the interplay of correlation effects
with topology, and it was argued to be a model topological Kondo
insulator~\cite{TopologicalKI-Dzero, TopologicalKI-PRB-Dzero,
  cubicTKI-arxiv-Alexandrov,SB_Lu}. It has also been argued that \SB
is an ideal topological insulator in which the conductance is
dominated by surface states at low temperature\cite{SB-trueTI-Wolgast,
 kim_robust_2012,SB-dopedtransport-Kim}.  Our prediction of \PB
with similar topological properties provides an alternative to these
intriguing studies, and may promote more experimental investigations
and theoretical understandings. In particular the larger energy scales
associated with a 5$f$ material should result in a more clear picture
of the surface bulk correspondence.  \PB is not only of high
scientific importance but may also have solid state properties
important for technological applications. The relatively small
hybridization gap and more importantly the topologically protected
metallic surface states of \PB with large Fermi surface pockets,
should result in an exceptionally high thermal conductivity for an
insulating materials. \PB has also a very high melting point
approximately 2200K. These are ideal solid state properties for a
nuclear fuel. Since B is a standard moderator in use in nuclear
reactors, \PB might be a model system in which to explore the
properties of topological nuclear fuels.

We acknowledge very useful discussions with S. Savrasov, X. Dai, J. Z. Zhao,
Z. P. Yin and N. Lanata. X. Deng and G. Kotliar are supported by
BES-DOE Grant DE-FG02-99ER45761, K. Haule is supported by the NSF DMR
0746395.


\begin{thebibliography}{50}%
\makeatletter
\providecommand \@ifxundefined [1]{%
 \@ifx{#1\undefined}
}%
\providecommand \@ifnum [1]{%
 \ifnum #1\expandafter \@firstoftwo
 \else \expandafter \@secondoftwo
 \fi
}%
\providecommand \@ifx [1]{%
 \ifx #1\expandafter \@firstoftwo
 \else \expandafter \@secondoftwo
 \fi
}%
\providecommand \natexlab [1]{#1}%
\providecommand \enquote  [1]{``#1''}%
\providecommand \bibnamefont  [1]{#1}%
\providecommand \bibfnamefont [1]{#1}%
\providecommand \citenamefont [1]{#1}%
\providecommand \href@noop [0]{\@secondoftwo}%
\providecommand \href [0]{\begingroup \@sanitize@url \@href}%
\providecommand \@href[1]{\@@startlink{#1}\@@href}%
\providecommand \@@href[1]{\endgroup#1\@@endlink}%
\providecommand \@sanitize@url [0]{\catcode `\\12\catcode `\$12\catcode
  `\&12\catcode `\#12\catcode `\^12\catcode `\_12\catcode `\%12\relax}%
\providecommand \@@startlink[1]{}%
\providecommand \@@endlink[0]{}%
\providecommand \url  [0]{\begingroup\@sanitize@url \@url }%
\providecommand \@url [1]{\endgroup\@href {#1}{\urlprefix }}%
\providecommand \urlprefix  [0]{URL }%
\providecommand \Eprint [0]{\href }%
\providecommand \doibase [0]{http://dx.doi.org/}%
\providecommand \selectlanguage [0]{\@gobble}%
\providecommand \bibinfo  [0]{\@secondoftwo}%
\providecommand \bibfield  [0]{\@secondoftwo}%
\providecommand \translation [1]{[#1]}%
\providecommand \BibitemOpen [0]{}%
\providecommand \bibitemStop [0]{}%
\providecommand \bibitemNoStop [0]{.\EOS\space}%
\providecommand \EOS [0]{\spacefactor3000\relax}%
\providecommand \BibitemShut  [1]{\csname bibitem#1\endcsname}%
\let\auto@bib@innerbib\@empty
\bibitem [{\citenamefont {Hecker}\ and\ \citenamefont
  {Timofeeva}(2000)}]{Pu-VolumeCollapse}%
  \BibitemOpen
  \bibfield  {author} {\bibinfo {author} {\bibfnamefont {S.~S.}\ \bibnamefont
  {Hecker}}\ and\ \bibinfo {author} {\bibfnamefont {L.~F.}\ \bibnamefont
  {Timofeeva}},\ }\href@noop {} {\bibfield  {journal} {\bibinfo  {journal} {Los
  Alamos Sci}\ }\textbf {\bibinfo {volume} {26}},\ \bibinfo {pages} {244}
  (\bibinfo {year} {2000})}\BibitemShut {NoStop}%
\bibitem [{\citenamefont {Sarrao}\ \emph {et~al.}(2002)\citenamefont {Sarrao},
  \citenamefont {Morales}, \citenamefont {Thompson}, \citenamefont {Scott},
  \citenamefont {Stewart}, \citenamefont {Wastin}, \citenamefont {Rebizant},
  \citenamefont {Boulet}, \citenamefont {Colineau},\ and\ \citenamefont
  {Lander}}]{sarrao_plutonium-based_2002}%
  \BibitemOpen
  \bibfield  {author} {\bibinfo {author} {\bibfnamefont {J.~L.}\ \bibnamefont
  {Sarrao}}, \bibinfo {author} {\bibfnamefont {L.~A.}\ \bibnamefont {Morales}},
  \bibinfo {author} {\bibfnamefont {J.~D.}\ \bibnamefont {Thompson}}, \bibinfo
  {author} {\bibfnamefont {B.~L.}\ \bibnamefont {Scott}}, \bibinfo {author}
  {\bibfnamefont {G.~R.}\ \bibnamefont {Stewart}}, \bibinfo {author}
  {\bibfnamefont {F.}~\bibnamefont {Wastin}}, \bibinfo {author} {\bibfnamefont
  {J.}~\bibnamefont {Rebizant}}, \bibinfo {author} {\bibfnamefont
  {P.}~\bibnamefont {Boulet}}, \bibinfo {author} {\bibfnamefont
  {E.}~\bibnamefont {Colineau}}, \ and\ \bibinfo {author} {\bibfnamefont
  {G.~H.}\ \bibnamefont {Lander}},\ }\href {\doibase 10.1038/nature01212}
  {\bibfield  {journal} {\bibinfo  {journal} {Nature}\ }\textbf {\bibinfo
  {volume} {420}},\ \bibinfo {pages} {297} (\bibinfo {year}
  {2002})}\BibitemShut {NoStop}%
\bibitem [{\citenamefont {Sarrao}\ and\ \citenamefont
  {Thompson}(2007)}]{SCinHF-review-Sarrao}%
  \BibitemOpen
  \bibfield  {author} {\bibinfo {author} {\bibfnamefont {J.~L.}\ \bibnamefont
  {Sarrao}}\ and\ \bibinfo {author} {\bibfnamefont {J.~D.}\ \bibnamefont
  {Thompson}},\ }\href {\doibase 10.1143/JPSJ.76.051013} {\bibfield  {journal}
  {\bibinfo  {journal} {Journal of the Physical Society of Japan}\ }\textbf
  {\bibinfo {volume} {76}},\ \bibinfo {pages} {051013} (\bibinfo {year}
  {2007})}\BibitemShut {NoStop}%
\bibitem [{\citenamefont {Qi}\ and\ \citenamefont {Zhang}(2011)}]{TI-RMP-Qi}%
  \BibitemOpen
  \bibfield  {author} {\bibinfo {author} {\bibfnamefont {X.-L.}\ \bibnamefont
  {Qi}}\ and\ \bibinfo {author} {\bibfnamefont {S.-C.}\ \bibnamefont {Zhang}},\
  }\href {\doibase 10.1103/RevModPhys.83.1057} {\bibfield  {journal} {\bibinfo
  {journal} {Rev. Mod. Phys.}\ }\textbf {\bibinfo {volume} {83}},\ \bibinfo
  {pages} {1057} (\bibinfo {year} {2011})}\BibitemShut {NoStop}%
\bibitem [{\citenamefont {Hasan}\ and\ \citenamefont
  {Kane}(2010)}]{TI-RMP-Hasan}%
  \BibitemOpen
  \bibfield  {author} {\bibinfo {author} {\bibfnamefont {M.~Z.}\ \bibnamefont
  {Hasan}}\ and\ \bibinfo {author} {\bibfnamefont {C.~L.}\ \bibnamefont
  {Kane}},\ }\href {\doibase 10.1103/RevModPhys.82.3045} {\bibfield  {journal}
  {\bibinfo  {journal} {Rev. Mod. Phys.}\ }\textbf {\bibinfo {volume} {82}},\
  \bibinfo {pages} {3045} (\bibinfo {year} {2010})}\BibitemShut {NoStop}%
\bibitem [{\citenamefont {Takimoto}(2011)}]{SB-Takimoto}%
  \BibitemOpen
  \bibfield  {author} {\bibinfo {author} {\bibfnamefont {T.}~\bibnamefont
  {Takimoto}},\ }\href {\doibase 10.1143/JPSJ.80.123710} {\bibfield  {journal}
  {\bibinfo  {journal} {Journal of the Physical Society of Japan}\ }\textbf
  {\bibinfo {volume} {80}},\ \bibinfo {pages} {123710} (\bibinfo {year}
  {2011})}\BibitemShut {NoStop}%
\bibitem [{\citenamefont {Dzero}\ \emph {et~al.}(2010)\citenamefont {Dzero},
  \citenamefont {Sun}, \citenamefont {Galitski},\ and\ \citenamefont
  {Coleman}}]{TopologicalKI-Dzero}%
  \BibitemOpen
  \bibfield  {author} {\bibinfo {author} {\bibfnamefont {M.}~\bibnamefont
  {Dzero}}, \bibinfo {author} {\bibfnamefont {K.}~\bibnamefont {Sun}}, \bibinfo
  {author} {\bibfnamefont {V.}~\bibnamefont {Galitski}}, \ and\ \bibinfo
  {author} {\bibfnamefont {P.}~\bibnamefont {Coleman}},\ }\href {\doibase
  10.1103/PhysRevLett.104.106408} {\bibfield  {journal} {\bibinfo  {journal}
  {Phys. Rev. Lett.}\ }\textbf {\bibinfo {volume} {104}},\ \bibinfo {pages}
  {106408} (\bibinfo {year} {2010})}\BibitemShut {NoStop}%
\bibitem [{\citenamefont {Dzero}\ \emph {et~al.}(2012)\citenamefont {Dzero},
  \citenamefont {Sun}, \citenamefont {Coleman},\ and\ \citenamefont
  {Galitski}}]{TopologicalKI-PRB-Dzero}%
  \BibitemOpen
  \bibfield  {author} {\bibinfo {author} {\bibfnamefont {M.}~\bibnamefont
  {Dzero}}, \bibinfo {author} {\bibfnamefont {K.}~\bibnamefont {Sun}}, \bibinfo
  {author} {\bibfnamefont {P.}~\bibnamefont {Coleman}}, \ and\ \bibinfo
  {author} {\bibfnamefont {V.}~\bibnamefont {Galitski}},\ }\href {\doibase
  10.1103/PhysRevB.85.045130} {\bibfield  {journal} {\bibinfo  {journal} {Phys.
  Rev. B}\ }\textbf {\bibinfo {volume} {85}},\ \bibinfo {pages} {045130}
  (\bibinfo {year} {2012})}\BibitemShut {NoStop}%
\bibitem [{\citenamefont {Lu}\ \emph {et~al.}(2013)\citenamefont {Lu},
  \citenamefont {Zhao}, \citenamefont {Weng}, \citenamefont {Fang},\ and\
  \citenamefont {Dai}}]{SB_Lu}%
  \BibitemOpen
  \bibfield  {author} {\bibinfo {author} {\bibfnamefont {F.}~\bibnamefont
  {Lu}}, \bibinfo {author} {\bibfnamefont {J.}~\bibnamefont {Zhao}}, \bibinfo
  {author} {\bibfnamefont {H.}~\bibnamefont {Weng}}, \bibinfo {author}
  {\bibfnamefont {Z.}~\bibnamefont {Fang}}, \ and\ \bibinfo {author}
  {\bibfnamefont {X.}~\bibnamefont {Dai}},\ }\href {\doibase
  10.1103/PhysRevLett.110.096401} {\bibfield  {journal} {\bibinfo  {journal}
  {Phys. Rev. Lett.}\ }\textbf {\bibinfo {volume} {110}},\ \bibinfo {pages}
  {096401} (\bibinfo {year} {2013})}\BibitemShut {NoStop}%
\bibitem [{\citenamefont {Alexandrov}\ \emph {et~al.}(2013)\citenamefont
  {Alexandrov}, \citenamefont {Dzero},\ and\ \citenamefont
  {Coleman}}]{cubicTKI-arxiv-Alexandrov}%
  \BibitemOpen
  \bibfield  {author} {\bibinfo {author} {\bibfnamefont {V.}~\bibnamefont
  {Alexandrov}}, \bibinfo {author} {\bibfnamefont {M.}~\bibnamefont {Dzero}}, \
  and\ \bibinfo {author} {\bibfnamefont {P.}~\bibnamefont {Coleman}},\
  }\href@noop {} {\  (\bibinfo {year} {2013})},\ \Eprint
  {http://arxiv.org/abs/1303.7224} {arXiv:1303.7224} \BibitemShut {NoStop}%
\bibitem [{\citenamefont {Miyazaki}\ \emph {et~al.}(2012)\citenamefont
  {Miyazaki}, \citenamefont {Hajiri}, \citenamefont {Ito}, \citenamefont
  {Kunii},\ and\ \citenamefont {Kimura}}]{Miyazaki-Momentum-2012}%
  \BibitemOpen
  \bibfield  {author} {\bibinfo {author} {\bibfnamefont {H.}~\bibnamefont
  {Miyazaki}}, \bibinfo {author} {\bibfnamefont {T.}~\bibnamefont {Hajiri}},
  \bibinfo {author} {\bibfnamefont {T.}~\bibnamefont {Ito}}, \bibinfo {author}
  {\bibfnamefont {S.}~\bibnamefont {Kunii}}, \ and\ \bibinfo {author}
  {\bibfnamefont {S.-i.}\ \bibnamefont {Kimura}},\ }\href {\doibase
  10.1103/PhysRevB.86.075105} {\bibfield  {journal} {\bibinfo  {journal} {Phys.
  Rev. B}\ }\textbf {\bibinfo {volume} {86}},\ \bibinfo {pages} {075105}
  (\bibinfo {year} {2012})}\BibitemShut {NoStop}%
\bibitem [{\citenamefont {Wolgast}\ \emph {et~al.}(2012)\citenamefont
  {Wolgast}, \citenamefont {Kurdak}, \citenamefont {Sun}, \citenamefont
  {Allen}, \citenamefont {Kim},\ and\ \citenamefont
  {Fisk}}]{SB-trueTI-Wolgast}%
  \BibitemOpen
  \bibfield  {author} {\bibinfo {author} {\bibfnamefont {S.}~\bibnamefont
  {Wolgast}}, \bibinfo {author} {\bibfnamefont {C.}~\bibnamefont {Kurdak}},
  \bibinfo {author} {\bibfnamefont {K.}~\bibnamefont {Sun}}, \bibinfo {author}
  {\bibfnamefont {J.~W.}\ \bibnamefont {Allen}}, \bibinfo {author}
  {\bibfnamefont {D.-J.}\ \bibnamefont {Kim}}, \ and\ \bibinfo {author}
  {\bibfnamefont {Z.}~\bibnamefont {Fisk}},\ }\href
  {http://arxiv.org/abs/1211.5104} {\  (\bibinfo {year} {2012})},\ \Eprint
  {http://arxiv.org/abs/1211.5104} {arXiv:1211.5104} \BibitemShut {NoStop}%
\bibitem [{\citenamefont {Zhang}\ \emph {et~al.}(2013)\citenamefont {Zhang},
  \citenamefont {Butch}, \citenamefont {Syers}, \citenamefont {Ziemak},
  \citenamefont {Greene},\ and\ \citenamefont
  {Paglione}}]{zhang_hybridization_2013}%
  \BibitemOpen
  \bibfield  {author} {\bibinfo {author} {\bibfnamefont {X.}~\bibnamefont
  {Zhang}}, \bibinfo {author} {\bibfnamefont {N.~P.}\ \bibnamefont {Butch}},
  \bibinfo {author} {\bibfnamefont {P.}~\bibnamefont {Syers}}, \bibinfo
  {author} {\bibfnamefont {S.}~\bibnamefont {Ziemak}}, \bibinfo {author}
  {\bibfnamefont {R.~L.}\ \bibnamefont {Greene}}, \ and\ \bibinfo {author}
  {\bibfnamefont {J.}~\bibnamefont {Paglione}},\ }\href {\doibase
  10.1103/PhysRevX.3.011011} {\bibfield  {journal} {\bibinfo  {journal} {Phys.
  Rev. X}\ }\textbf {\bibinfo {volume} {3}},\ \bibinfo {pages} {011011}
  (\bibinfo {year} {2013})}\BibitemShut {NoStop}%
\bibitem [{\citenamefont {Kim}\ \emph {et~al.}(2012)\citenamefont {Kim},
  \citenamefont {Thomas}, \citenamefont {Grant}, \citenamefont {Botimer},
  \citenamefont {Fisk},\ and\ \citenamefont {Xia}}]{kim_robust_2012}%
  \BibitemOpen
  \bibfield  {author} {\bibinfo {author} {\bibfnamefont {D.~J.}\ \bibnamefont
  {Kim}}, \bibinfo {author} {\bibfnamefont {S.}~\bibnamefont {Thomas}},
  \bibinfo {author} {\bibfnamefont {T.}~\bibnamefont {Grant}}, \bibinfo
  {author} {\bibfnamefont {J.}~\bibnamefont {Botimer}}, \bibinfo {author}
  {\bibfnamefont {Z.}~\bibnamefont {Fisk}}, \ and\ \bibinfo {author}
  {\bibfnamefont {J.}~\bibnamefont {Xia}},\ }\href
  {http://arxiv.org/abs/1211.6769} {\  (\bibinfo {year} {2012})},\ \Eprint
  {http://arxiv.org/abs/1211.6769} {arXiv:1211.6769} \BibitemShut {NoStop}%
\bibitem [{\citenamefont {Xu}\ \emph {et~al.}(2013)\citenamefont {Xu},
  \citenamefont {Shi}, \citenamefont {Biswas}, \citenamefont {Matt},
  \citenamefont {Dhaka}, \citenamefont {Huang}, \citenamefont {Plumb},
  \citenamefont {Radovic}, \citenamefont {Dil}, \citenamefont {Pomjakushina},
  \citenamefont {Amato}, \citenamefont {Salman}, \citenamefont {Paul},
  \citenamefont {Mesot}, \citenamefont {Ding},\ and\ \citenamefont
  {Shi}}]{xu_surface_2013}%
  \BibitemOpen
  \bibfield  {author} {\bibinfo {author} {\bibfnamefont {N.}~\bibnamefont
  {Xu}}, \bibinfo {author} {\bibfnamefont {X.}~\bibnamefont {Shi}}, \bibinfo
  {author} {\bibfnamefont {P.~K.}\ \bibnamefont {Biswas}}, \bibinfo {author}
  {\bibfnamefont {C.~E.}\ \bibnamefont {Matt}}, \bibinfo {author}
  {\bibfnamefont {R.~S.}\ \bibnamefont {Dhaka}}, \bibinfo {author}
  {\bibfnamefont {Y.}~\bibnamefont {Huang}}, \bibinfo {author} {\bibfnamefont
  {N.~C.}\ \bibnamefont {Plumb}}, \bibinfo {author} {\bibfnamefont
  {M.}~\bibnamefont {Radovic}}, \bibinfo {author} {\bibfnamefont {J.~H.}\
  \bibnamefont {Dil}}, \bibinfo {author} {\bibfnamefont {E.}~\bibnamefont
  {Pomjakushina}}, \bibinfo {author} {\bibfnamefont {A.}~\bibnamefont {Amato}},
  \bibinfo {author} {\bibfnamefont {Z.}~\bibnamefont {Salman}}, \bibinfo
  {author} {\bibfnamefont {D.~M.}\ \bibnamefont {Paul}}, \bibinfo {author}
  {\bibfnamefont {J.}~\bibnamefont {Mesot}}, \bibinfo {author} {\bibfnamefont
  {H.}~\bibnamefont {Ding}}, \ and\ \bibinfo {author} {\bibfnamefont
  {M.}~\bibnamefont {Shi}},\ }\href {http://arxiv.org/abs/1306.3678} {\
  (\bibinfo {year} {2013})},\ \Eprint {http://arxiv.org/abs/1306.3678}
  {arXiv:1306.3678} \BibitemShut {NoStop}%
\bibitem [{\citenamefont {Neupane}\ \emph {et~al.}(2013)\citenamefont
  {Neupane}, \citenamefont {Alidoust}, \citenamefont {Xu}, \citenamefont
  {Kondo}, \citenamefont {Kim}, \citenamefont {Liu}, \citenamefont
  {Belopolski}, \citenamefont {Chang}, \citenamefont {Jeng}, \citenamefont
  {Durakiewicz}, \citenamefont {Balicas}, \citenamefont {Lin}, \citenamefont
  {Bansil}, \citenamefont {Shin}, \citenamefont {Fisk},\ and\ \citenamefont
  {Hasan}}]{SB-ARPES-Neupane}%
  \BibitemOpen
  \bibfield  {author} {\bibinfo {author} {\bibfnamefont {M.}~\bibnamefont
  {Neupane}}, \bibinfo {author} {\bibfnamefont {N.}~\bibnamefont {Alidoust}},
  \bibinfo {author} {\bibfnamefont {S.-Y.}\ \bibnamefont {Xu}}, \bibinfo
  {author} {\bibfnamefont {T.}~\bibnamefont {Kondo}}, \bibinfo {author}
  {\bibfnamefont {D.-J.}\ \bibnamefont {Kim}}, \bibinfo {author} {\bibfnamefont
  {C.}~\bibnamefont {Liu}}, \bibinfo {author} {\bibfnamefont {I.}~\bibnamefont
  {Belopolski}}, \bibinfo {author} {\bibfnamefont {T.-R.}\ \bibnamefont
  {Chang}}, \bibinfo {author} {\bibfnamefont {H.-T.}\ \bibnamefont {Jeng}},
  \bibinfo {author} {\bibfnamefont {T.}~\bibnamefont {Durakiewicz}}, \bibinfo
  {author} {\bibfnamefont {L.}~\bibnamefont {Balicas}}, \bibinfo {author}
  {\bibfnamefont {H.}~\bibnamefont {Lin}}, \bibinfo {author} {\bibfnamefont
  {A.}~\bibnamefont {Bansil}}, \bibinfo {author} {\bibfnamefont
  {S.}~\bibnamefont {Shin}}, \bibinfo {author} {\bibfnamefont {Z.}~\bibnamefont
  {Fisk}}, \ and\ \bibinfo {author} {\bibfnamefont {M.~Z.}\ \bibnamefont
  {Hasan}},\ }\href {http://arxiv.org/abs/1306.4634} {\  (\bibinfo {year}
  {2013})},\ \Eprint {http://arxiv.org/abs/1306.4634} {arXiv:1306.4634}
  \BibitemShut {NoStop}%
\bibitem [{\citenamefont {Li}\ \emph {et~al.}(2013)\citenamefont {Li},
  \citenamefont {Xiang}, \citenamefont {Yu}, \citenamefont {Asaba},
  \citenamefont {Lawson}, \citenamefont {Cai}, \citenamefont {Tinsman},
  \citenamefont {Berkley}, \citenamefont {Wolgast}, \citenamefont {Eo},
  \citenamefont {Kim}, \citenamefont {Kurdak}, \citenamefont {Allen},
  \citenamefont {Sun}, \citenamefont {Chen}, \citenamefont {Wang},
  \citenamefont {Fisk},\ and\ \citenamefont {Li}}]{li_quantum_2013}%
  \BibitemOpen
  \bibfield  {author} {\bibinfo {author} {\bibfnamefont {G.}~\bibnamefont
  {Li}}, \bibinfo {author} {\bibfnamefont {Z.}~\bibnamefont {Xiang}}, \bibinfo
  {author} {\bibfnamefont {F.}~\bibnamefont {Yu}}, \bibinfo {author}
  {\bibfnamefont {T.}~\bibnamefont {Asaba}}, \bibinfo {author} {\bibfnamefont
  {B.}~\bibnamefont {Lawson}}, \bibinfo {author} {\bibfnamefont
  {P.}~\bibnamefont {Cai}}, \bibinfo {author} {\bibfnamefont {C.}~\bibnamefont
  {Tinsman}}, \bibinfo {author} {\bibfnamefont {A.}~\bibnamefont {Berkley}},
  \bibinfo {author} {\bibfnamefont {S.}~\bibnamefont {Wolgast}}, \bibinfo
  {author} {\bibfnamefont {Y.~S.}\ \bibnamefont {Eo}}, \bibinfo {author}
  {\bibfnamefont {D.-J.}\ \bibnamefont {Kim}}, \bibinfo {author} {\bibfnamefont
  {C.}~\bibnamefont {Kurdak}}, \bibinfo {author} {\bibfnamefont {J.~W.}\
  \bibnamefont {Allen}}, \bibinfo {author} {\bibfnamefont {K.}~\bibnamefont
  {Sun}}, \bibinfo {author} {\bibfnamefont {X.~H.}\ \bibnamefont {Chen}},
  \bibinfo {author} {\bibfnamefont {Y.~Y.}\ \bibnamefont {Wang}}, \bibinfo
  {author} {\bibfnamefont {Z.}~\bibnamefont {Fisk}}, \ and\ \bibinfo {author}
  {\bibfnamefont {L.}~\bibnamefont {Li}},\ }\href
  {http://arxiv.org/abs/1306.5221} {\  (\bibinfo {year} {2013})},\ \Eprint
  {http://arxiv.org/abs/1306.5221} {arXiv:1306.5221} \BibitemShut {NoStop}%
\bibitem [{\citenamefont {Jiang}\ \emph {et~al.}(2013)\citenamefont {Jiang},
  \citenamefont {Li}, \citenamefont {Zhang}, \citenamefont {Sun}, \citenamefont
  {Chen}, \citenamefont {Ye}, \citenamefont {Xu}, \citenamefont {Ge},
  \citenamefont {Tan}, \citenamefont {Niu}, \citenamefont {Xia}, \citenamefont
  {Xie}, \citenamefont {Li}, \citenamefont {Chen}, \citenamefont {Wen},\ and\
  \citenamefont {Feng}}]{SB-ARPES-Jiang}%
  \BibitemOpen
  \bibfield  {author} {\bibinfo {author} {\bibfnamefont {J.}~\bibnamefont
  {Jiang}}, \bibinfo {author} {\bibfnamefont {S.}~\bibnamefont {Li}}, \bibinfo
  {author} {\bibfnamefont {T.}~\bibnamefont {Zhang}}, \bibinfo {author}
  {\bibfnamefont {Z.}~\bibnamefont {Sun}}, \bibinfo {author} {\bibfnamefont
  {F.}~\bibnamefont {Chen}}, \bibinfo {author} {\bibfnamefont {Z.~R.}\
  \bibnamefont {Ye}}, \bibinfo {author} {\bibfnamefont {M.}~\bibnamefont {Xu}},
  \bibinfo {author} {\bibfnamefont {Q.~Q.}\ \bibnamefont {Ge}}, \bibinfo
  {author} {\bibfnamefont {S.~Y.}\ \bibnamefont {Tan}}, \bibinfo {author}
  {\bibfnamefont {X.~H.}\ \bibnamefont {Niu}}, \bibinfo {author} {\bibfnamefont
  {M.}~\bibnamefont {Xia}}, \bibinfo {author} {\bibfnamefont {B.~P.}\
  \bibnamefont {Xie}}, \bibinfo {author} {\bibfnamefont {Y.~F.}\ \bibnamefont
  {Li}}, \bibinfo {author} {\bibfnamefont {X.~H.}\ \bibnamefont {Chen}},
  \bibinfo {author} {\bibfnamefont {H.~H.}\ \bibnamefont {Wen}}, \ and\
  \bibinfo {author} {\bibfnamefont {D.~L.}\ \bibnamefont {Feng}},\ }\href@noop
  {} {\  (\bibinfo {year} {2013})},\ \Eprint {http://arxiv.org/abs/1306.5664}
  {arXiv:1306.5664} \BibitemShut {NoStop}%
\bibitem [{\citenamefont {Kim}\ \emph {et~al.}(2013)\citenamefont {Kim},
  \citenamefont {Xia},\ and\ \citenamefont {Fisk}}]{SB-dopedtransport-Kim}%
  \BibitemOpen
  \bibfield  {author} {\bibinfo {author} {\bibfnamefont {D.~J.}\ \bibnamefont
  {Kim}}, \bibinfo {author} {\bibfnamefont {J.}~\bibnamefont {Xia}}, \ and\
  \bibinfo {author} {\bibfnamefont {Z.}~\bibnamefont {Fisk}},\ }\href
  {http://arxiv.org/abs/1307.0448} {\  (\bibinfo {year} {2013})},\ \Eprint
  {http://arxiv.org/abs/1307.0448} {arXiv:1307.0448} \BibitemShut {NoStop}%
\bibitem [{\citenamefont {Thomas}\ \emph {et~al.}(2013)\citenamefont {Thomas},
  \citenamefont {Kim}, \citenamefont {Chung}, \citenamefont {Grant},
  \citenamefont {Fisk},\ and\ \citenamefont {Xia}}]{thomas_weak_2013}%
  \BibitemOpen
  \bibfield  {author} {\bibinfo {author} {\bibfnamefont {S.}~\bibnamefont
  {Thomas}}, \bibinfo {author} {\bibfnamefont {D.~J.}\ \bibnamefont {Kim}},
  \bibinfo {author} {\bibfnamefont {S.~B.}\ \bibnamefont {Chung}}, \bibinfo
  {author} {\bibfnamefont {T.}~\bibnamefont {Grant}}, \bibinfo {author}
  {\bibfnamefont {Z.}~\bibnamefont {Fisk}}, \ and\ \bibinfo {author}
  {\bibfnamefont {J.}~\bibnamefont {Xia}},\ }\href
  {http://arxiv.org/abs/1307.4133} {\  (\bibinfo {year} {2013})},\ \Eprint
  {http://arxiv.org/abs/1307.4133} {arXiv:1307.4133} \BibitemShut {NoStop}%
\bibitem [{\citenamefont {Frantzeskakis}\ \emph {et~al.}(2013)\citenamefont
  {Frantzeskakis}, \citenamefont {de~Jong}, \citenamefont {Zwartsenberg},
  \citenamefont {Huang}, \citenamefont {Pan}, \citenamefont {Zhang},
  \citenamefont {Zhang}, \citenamefont {Zhang}, \citenamefont {Bao},
  \citenamefont {Tegus}, \citenamefont {Varykhalov}, \citenamefont
  {de~Visser},\ and\ \citenamefont {Golden}}]{frantzeskakis_kondo_2013}%
  \BibitemOpen
  \bibfield  {author} {\bibinfo {author} {\bibfnamefont {E.}~\bibnamefont
  {Frantzeskakis}}, \bibinfo {author} {\bibfnamefont {N.}~\bibnamefont
  {de~Jong}}, \bibinfo {author} {\bibfnamefont {B.}~\bibnamefont
  {Zwartsenberg}}, \bibinfo {author} {\bibfnamefont {Y.~K.}\ \bibnamefont
  {Huang}}, \bibinfo {author} {\bibfnamefont {Y.}~\bibnamefont {Pan}}, \bibinfo
  {author} {\bibfnamefont {X.}~\bibnamefont {Zhang}}, \bibinfo {author}
  {\bibfnamefont {J.~X.}\ \bibnamefont {Zhang}}, \bibinfo {author}
  {\bibfnamefont {F.~X.}\ \bibnamefont {Zhang}}, \bibinfo {author}
  {\bibfnamefont {L.~H.}\ \bibnamefont {Bao}}, \bibinfo {author} {\bibfnamefont
  {O.}~\bibnamefont {Tegus}}, \bibinfo {author} {\bibfnamefont
  {A.}~\bibnamefont {Varykhalov}}, \bibinfo {author} {\bibfnamefont
  {A.}~\bibnamefont {de~Visser}}, \ and\ \bibinfo {author} {\bibfnamefont
  {M.~S.}\ \bibnamefont {Golden}},\ }\href {http://arxiv.org/abs/1308.0151} {\
  (\bibinfo {year} {2013})},\ \Eprint {http://arxiv.org/abs/1308.0151}
  {arXiv:1308.0151} \BibitemShut {NoStop}%
\bibitem [{\citenamefont {Yee}\ \emph {et~al.}(2013)\citenamefont {Yee},
  \citenamefont {He}, \citenamefont {Soumyanarayanan}, \citenamefont {Kim},
  \citenamefont {Fisk},\ and\ \citenamefont {Hoffman}}]{yee_imaging_2013}%
  \BibitemOpen
  \bibfield  {author} {\bibinfo {author} {\bibfnamefont {M.~M.}\ \bibnamefont
  {Yee}}, \bibinfo {author} {\bibfnamefont {Y.}~\bibnamefont {He}}, \bibinfo
  {author} {\bibfnamefont {A.}~\bibnamefont {Soumyanarayanan}}, \bibinfo
  {author} {\bibfnamefont {D.-J.}\ \bibnamefont {Kim}}, \bibinfo {author}
  {\bibfnamefont {Z.}~\bibnamefont {Fisk}}, \ and\ \bibinfo {author}
  {\bibfnamefont {J.~E.}\ \bibnamefont {Hoffman}},\ }\href
  {http://arxiv.org/abs/1308.1085} {\  (\bibinfo {year} {2013})},\ \Eprint
  {http://arxiv.org/abs/1308.1085} {arXiv:1308.1085} \BibitemShut {NoStop}%
\bibitem [{\citenamefont {Rogl}\ and\ \citenamefont
  {Potter}(1997)}]{B-Pu-review}%
  \BibitemOpen
  \bibfield  {author} {\bibinfo {author} {\bibfnamefont {P.}~\bibnamefont
  {Rogl}}\ and\ \bibinfo {author} {\bibfnamefont {P.}~\bibnamefont {Potter}},\
  }\href {\doibase 10.1007/BF02647703} {\bibfield  {journal} {\bibinfo
  {journal} {Journal of Phase Equilibria}\ }\textbf {\bibinfo {volume} {18}},\
  \bibinfo {pages} {467} (\bibinfo {year} {1997})}\BibitemShut {NoStop}%
\bibitem [{\citenamefont {Kane}\ and\ \citenamefont {Mele}(2005)}]{TI-Z2-Kane}%
  \BibitemOpen
  \bibfield  {author} {\bibinfo {author} {\bibfnamefont {C.~L.}\ \bibnamefont
  {Kane}}\ and\ \bibinfo {author} {\bibfnamefont {E.~J.}\ \bibnamefont
  {Mele}},\ }\href {\doibase 10.1103/PhysRevLett.95.146802} {\bibfield
  {journal} {\bibinfo  {journal} {Phys. Rev. Lett.}\ }\textbf {\bibinfo
  {volume} {95}},\ \bibinfo {pages} {146802} (\bibinfo {year}
  {2005})}\BibitemShut {NoStop}%
\bibitem [{\citenamefont {Fu}\ and\ \citenamefont
  {Kane}(2007)}]{TI-inversionsymmetry-Fu}%
  \BibitemOpen
  \bibfield  {author} {\bibinfo {author} {\bibfnamefont {L.}~\bibnamefont
  {Fu}}\ and\ \bibinfo {author} {\bibfnamefont {C.~L.}\ \bibnamefont {Kane}},\
  }\href {\doibase 10.1103/PhysRevB.76.045302} {\bibfield  {journal} {\bibinfo
  {journal} {Phys. Rev. B}\ }\textbf {\bibinfo {volume} {76}},\ \bibinfo
  {pages} {045302} (\bibinfo {year} {2007})}\BibitemShut {NoStop}%
\bibitem [{\citenamefont {Moore}\ and\ \citenamefont
  {Balents}(2007)}]{TI-Z2-Moore}%
  \BibitemOpen
  \bibfield  {author} {\bibinfo {author} {\bibfnamefont {J.~E.}\ \bibnamefont
  {Moore}}\ and\ \bibinfo {author} {\bibfnamefont {L.}~\bibnamefont
  {Balents}},\ }\href {\doibase 10.1103/PhysRevB.75.121306} {\bibfield
  {journal} {\bibinfo  {journal} {Phys. Rev. B}\ }\textbf {\bibinfo {volume}
  {75}},\ \bibinfo {pages} {121306} (\bibinfo {year} {2007})}\BibitemShut
  {NoStop}%
\bibitem [{\citenamefont {Qi}\ \emph {et~al.}(2008)\citenamefont {Qi},
  \citenamefont {Hughes},\ and\ \citenamefont {Zhang}}]{TI-fieldtheory-Qi}%
  \BibitemOpen
  \bibfield  {author} {\bibinfo {author} {\bibfnamefont {X.-L.}\ \bibnamefont
  {Qi}}, \bibinfo {author} {\bibfnamefont {T.~L.}\ \bibnamefont {Hughes}}, \
  and\ \bibinfo {author} {\bibfnamefont {S.-C.}\ \bibnamefont {Zhang}},\ }\href
  {\doibase 10.1103/PhysRevB.78.195424} {\bibfield  {journal} {\bibinfo
  {journal} {Phys. Rev. B}\ }\textbf {\bibinfo {volume} {78}},\ \bibinfo
  {pages} {195424} (\bibinfo {year} {2008})}\BibitemShut {NoStop}%
\bibitem [{\citenamefont {Wang}\ and\ \citenamefont
  {Zhang}(2012)}]{TI-Z2interacting-Wang}%
  \BibitemOpen
  \bibfield  {author} {\bibinfo {author} {\bibfnamefont {Z.}~\bibnamefont
  {Wang}}\ and\ \bibinfo {author} {\bibfnamefont {S.-C.}\ \bibnamefont
  {Zhang}},\ }\href {\doibase 10.1103/PhysRevX.2.031008} {\bibfield  {journal}
  {\bibinfo  {journal} {Phys. Rev. X}\ }\textbf {\bibinfo {volume} {2}},\
  \bibinfo {pages} {031008} (\bibinfo {year} {2012})}\BibitemShut {NoStop}%
\bibitem [{\citenamefont {Bernevig}\ \emph {et~al.}(2006)\citenamefont
  {Bernevig}, \citenamefont {Hughes},\ and\ \citenamefont
  {Zhang}}]{HgTe-science-Bernevig}%
  \BibitemOpen
  \bibfield  {author} {\bibinfo {author} {\bibfnamefont {B.~A.}\ \bibnamefont
  {Bernevig}}, \bibinfo {author} {\bibfnamefont {T.~L.}\ \bibnamefont
  {Hughes}}, \ and\ \bibinfo {author} {\bibfnamefont {S.-C.}\ \bibnamefont
  {Zhang}},\ }\href {\doibase 10.1126/science.1133734} {\bibfield  {journal}
  {\bibinfo  {journal} {Science}\ }\textbf {\bibinfo {volume} {314}},\ \bibinfo
  {pages} {1757} (\bibinfo {year} {2006})}\BibitemShut {NoStop}%
\bibitem [{\citenamefont {Zhang}\ \emph {et~al.}(2009)\citenamefont {Zhang},
  \citenamefont {Liu}, \citenamefont {Qi}, \citenamefont {Dai}, \citenamefont
  {Fang},\ and\ \citenamefont {Zhang}}]{Bi2Se3-NP-Zhang}%
  \BibitemOpen
  \bibfield  {author} {\bibinfo {author} {\bibfnamefont {H.}~\bibnamefont
  {Zhang}}, \bibinfo {author} {\bibfnamefont {C.-X.}\ \bibnamefont {Liu}},
  \bibinfo {author} {\bibfnamefont {X.-L.}\ \bibnamefont {Qi}}, \bibinfo
  {author} {\bibfnamefont {X.}~\bibnamefont {Dai}}, \bibinfo {author}
  {\bibfnamefont {Z.}~\bibnamefont {Fang}}, \ and\ \bibinfo {author}
  {\bibfnamefont {S.-C.}\ \bibnamefont {Zhang}},\ }\href {\doibase
  10.1038/nphys1270} {\bibfield  {journal} {\bibinfo  {journal} {Nature
  Physics}\ }\textbf {\bibinfo {volume} {5}},\ \bibinfo {pages} {438} (\bibinfo
  {year} {2009})}\BibitemShut {NoStop}%
\bibitem [{\citenamefont {Fu}\ \emph {et~al.}(2007)\citenamefont {Fu},
  \citenamefont {Kane},\ and\ \citenamefont {Mele}}]{3DTI-Fu}%
  \BibitemOpen
  \bibfield  {author} {\bibinfo {author} {\bibfnamefont {L.}~\bibnamefont
  {Fu}}, \bibinfo {author} {\bibfnamefont {C.~L.}\ \bibnamefont {Kane}}, \ and\
  \bibinfo {author} {\bibfnamefont {E.~J.}\ \bibnamefont {Mele}},\ }\href
  {\doibase 10.1103/PhysRevLett.98.106803} {\bibfield  {journal} {\bibinfo
  {journal} {Phys. Rev. Lett.}\ }\textbf {\bibinfo {volume} {98}},\ \bibinfo
  {pages} {106803} (\bibinfo {year} {2007})}\BibitemShut {NoStop}%
\bibitem [{\citenamefont {Zhang}\ and\ \citenamefont
  {Zhang}(2013)}]{TI-Firstprinciple-Zhang}%
  \BibitemOpen
  \bibfield  {author} {\bibinfo {author} {\bibfnamefont {H.}~\bibnamefont
  {Zhang}}\ and\ \bibinfo {author} {\bibfnamefont {S.-C.}\ \bibnamefont
  {Zhang}},\ }\href {\doibase 10.1002/pssr.201390006} {\bibfield  {journal}
  {\bibinfo  {journal} {physica status solidi (RRL) – Rapid Research
  Letters}\ }\textbf {\bibinfo {volume} {7}},\ \bibinfo {pages} {n/a} (\bibinfo
  {year} {2013})}\BibitemShut {NoStop}%
\bibitem [{\citenamefont {Wang}\ \emph {et~al.}(2012)\citenamefont {Wang},
  \citenamefont {Qi},\ and\ \citenamefont {Zhang}}]{TI-Z2interactingPRB-Wang}%
  \BibitemOpen
  \bibfield  {author} {\bibinfo {author} {\bibfnamefont {Z.}~\bibnamefont
  {Wang}}, \bibinfo {author} {\bibfnamefont {X.-L.}\ \bibnamefont {Qi}}, \ and\
  \bibinfo {author} {\bibfnamefont {S.-C.}\ \bibnamefont {Zhang}},\ }\href
  {\doibase 10.1103/PhysRevB.85.165126} {\bibfield  {journal} {\bibinfo
  {journal} {Phys. Rev. B}\ }\textbf {\bibinfo {volume} {85}},\ \bibinfo
  {pages} {165126} (\bibinfo {year} {2012})}\BibitemShut {NoStop}%
\bibitem [{\citenamefont {Qi}\ \emph {et~al.}(2009)\citenamefont {Qi},
  \citenamefont {Li}, \citenamefont {Zang},\ and\ \citenamefont
  {Zhang}}]{TI-monopole-Qi}%
  \BibitemOpen
  \bibfield  {author} {\bibinfo {author} {\bibfnamefont {X.-L.}\ \bibnamefont
  {Qi}}, \bibinfo {author} {\bibfnamefont {R.}~\bibnamefont {Li}}, \bibinfo
  {author} {\bibfnamefont {J.}~\bibnamefont {Zang}}, \ and\ \bibinfo {author}
  {\bibfnamefont {S.-C.}\ \bibnamefont {Zhang}},\ }\href {\doibase
  10.1126/science.1167747} {\bibfield  {journal} {\bibinfo  {journal}
  {Science}\ }\textbf {\bibinfo {volume} {323}},\ \bibinfo {pages} {1184}
  (\bibinfo {year} {2009})}\BibitemShut {NoStop}%
\bibitem [{\citenamefont {Zhang}\ \emph {et~al.}(2012)\citenamefont {Zhang},
  \citenamefont {Zhang}, \citenamefont {Wang}, \citenamefont {Felser},\ and\
  \citenamefont {Zhang}}]{TI-Actinides-Zhang}%
  \BibitemOpen
  \bibfield  {author} {\bibinfo {author} {\bibfnamefont {X.}~\bibnamefont
  {Zhang}}, \bibinfo {author} {\bibfnamefont {H.}~\bibnamefont {Zhang}},
  \bibinfo {author} {\bibfnamefont {J.}~\bibnamefont {Wang}}, \bibinfo {author}
  {\bibfnamefont {C.}~\bibnamefont {Felser}}, \ and\ \bibinfo {author}
  {\bibfnamefont {S.-C.}\ \bibnamefont {Zhang}},\ }\href {\doibase
  10.1126/science.1216184} {\bibfield  {journal} {\bibinfo  {journal}
  {Science}\ }\textbf {\bibinfo {volume} {335}},\ \bibinfo {pages} {1464}
  (\bibinfo {year} {2012})}\BibitemShut {NoStop}%
\bibitem [{\citenamefont {Kotliar}\ \emph {et~al.}(2006)\citenamefont
  {Kotliar}, \citenamefont {Savrasov}, \citenamefont {Haule}, \citenamefont
  {Oudovenko}, \citenamefont {Parcollet},\ and\ \citenamefont
  {Marianetti}}]{LDADMFT-RMP-Gabi}%
  \BibitemOpen
  \bibfield  {author} {\bibinfo {author} {\bibfnamefont {G.}~\bibnamefont
  {Kotliar}}, \bibinfo {author} {\bibfnamefont {S.~Y.}\ \bibnamefont
  {Savrasov}}, \bibinfo {author} {\bibfnamefont {K.}~\bibnamefont {Haule}},
  \bibinfo {author} {\bibfnamefont {V.~S.}\ \bibnamefont {Oudovenko}}, \bibinfo
  {author} {\bibfnamefont {O.}~\bibnamefont {Parcollet}}, \ and\ \bibinfo
  {author} {\bibfnamefont {C.~A.}\ \bibnamefont {Marianetti}},\ }\href
  {\doibase 10.1103/RevModPhys.78.865} {\bibfield  {journal} {\bibinfo
  {journal} {Rev. Mod. Phys.}\ }\textbf {\bibinfo {volume} {78}},\ \bibinfo
  {pages} {865} (\bibinfo {year} {2006})}\BibitemShut {NoStop}%
\bibitem [{\citenamefont {Held}(2007)}]{LDADMFT-review-Held}%
  \BibitemOpen
  \bibfield  {author} {\bibinfo {author} {\bibfnamefont {K.}~\bibnamefont
  {Held}},\ }\href {\doibase 10.1080/00018730701619647} {\bibfield  {journal}
  {\bibinfo  {journal} {Advances in Physics}\ }\textbf {\bibinfo {volume}
  {56}},\ \bibinfo {pages} {829} (\bibinfo {year} {2007})}\BibitemShut
  {NoStop}%
\bibitem [{\citenamefont {Haule}\ \emph {et~al.}(2010)\citenamefont {Haule},
  \citenamefont {Yee},\ and\ \citenamefont {Kim}}]{LDADMFT_Haule}%
  \BibitemOpen
  \bibfield  {author} {\bibinfo {author} {\bibfnamefont {K.}~\bibnamefont
  {Haule}}, \bibinfo {author} {\bibfnamefont {C.-H.}\ \bibnamefont {Yee}}, \
  and\ \bibinfo {author} {\bibfnamefont {K.}~\bibnamefont {Kim}},\ }\href
  {\doibase 10.1103/PhysRevB.81.195107} {\bibfield  {journal} {\bibinfo
  {journal} {Phys. Rev. B}\ }\textbf {\bibinfo {volume} {81}},\ \bibinfo
  {pages} {195107} (\bibinfo {year} {2010})}\BibitemShut {NoStop}%
\bibitem [{\citenamefont {Blaha}\ \emph {et~al.}(2001)\citenamefont {Blaha},
  \citenamefont {Schwarz}, \citenamefont {Madsen}, \citenamefont {Kvasnicka},\
  and\ \citenamefont {Luitz}}]{wien2k}%
  \BibitemOpen
  \bibfield  {author} {\bibinfo {author} {\bibfnamefont {P.}~\bibnamefont
  {Blaha}}, \bibinfo {author} {\bibfnamefont {K.}~\bibnamefont {Schwarz}},
  \bibinfo {author} {\bibfnamefont {G.~K.~H.}\ \bibnamefont {Madsen}}, \bibinfo
  {author} {\bibfnamefont {D.}~\bibnamefont {Kvasnicka}}, \ and\ \bibinfo
  {author} {\bibfnamefont {J.}~\bibnamefont {Luitz}},\ }\href@noop {} {\emph
  {\bibinfo {title} {{WIEN2K}, {A}n {A}ugmented {P}lane {W}ave + {L}ocal
  {O}rbitals {P}rogram for {C}alculating {C}rystal {P}roperties}}}\ (\bibinfo
  {publisher} {{K}arlheinz Schwarz, Techn. Universit\"{a}t Wien, Austria},\
  \bibinfo {year} {2001})\BibitemShut {NoStop}%
\bibitem [{\citenamefont {Shim}\ \emph {et~al.}(2007)\citenamefont {Shim},
  \citenamefont {Haule},\ and\ \citenamefont {Kotliar}}]{Pu-valence-Shim}%
  \BibitemOpen
  \bibfield  {author} {\bibinfo {author} {\bibfnamefont {J.~H.}\ \bibnamefont
  {Shim}}, \bibinfo {author} {\bibfnamefont {K.}~\bibnamefont {Haule}}, \ and\
  \bibinfo {author} {\bibfnamefont {G.}~\bibnamefont {Kotliar}},\ }\href
  {\doibase 10.1038/nature05647} {\bibfield  {journal} {\bibinfo  {journal}
  {Nature}\ }\textbf {\bibinfo {volume} {446}},\ \bibinfo {pages} {513}
  (\bibinfo {year} {2007})}\BibitemShut {NoStop}%
\bibitem [{\citenamefont {Cowan}(1981)}]{cowan}%
  \BibitemOpen
  \bibfield  {author} {\bibinfo {author} {\bibfnamefont {R.~D.}\ \bibnamefont
  {Cowan}},\ }\href@noop {} {\emph {\bibinfo {title} {The Theory of Atomic
  Structure and Spectra}}}\ (\bibinfo  {publisher} {Univ. California Press},\
  \bibinfo {address} {Berkeley},\ \bibinfo {year} {1981})\BibitemShut {NoStop}%
\bibitem [{\citenamefont {Haule}(2007)}]{CTQMC_Haule}%
  \BibitemOpen
  \bibfield  {author} {\bibinfo {author} {\bibfnamefont {K.}~\bibnamefont
  {Haule}},\ }\href {\doibase 10.1103/PhysRevB.75.155113} {\bibfield  {journal}
  {\bibinfo  {journal} {Phys. Rev. B}\ }\textbf {\bibinfo {volume} {75}},\
  \bibinfo {pages} {155113} (\bibinfo {year} {2007})}\BibitemShut {NoStop}%
\bibitem [{\citenamefont {Werner}\ \emph {et~al.}(2006)\citenamefont {Werner},
  \citenamefont {Comanac}, \citenamefont {de' Medici}, \citenamefont {Troyer},\
  and\ \citenamefont {Millis}}]{CTQMC_Werner}%
  \BibitemOpen
  \bibfield  {author} {\bibinfo {author} {\bibfnamefont {P.}~\bibnamefont
  {Werner}}, \bibinfo {author} {\bibfnamefont {A.}~\bibnamefont {Comanac}},
  \bibinfo {author} {\bibfnamefont {L.}~\bibnamefont {de' Medici}}, \bibinfo
  {author} {\bibfnamefont {M.}~\bibnamefont {Troyer}}, \ and\ \bibinfo {author}
  {\bibfnamefont {A.~J.}\ \bibnamefont {Millis}},\ }\href {\doibase
  10.1103/PhysRevLett.97.076405} {\bibfield  {journal} {\bibinfo  {journal}
  {Phys. Rev. Lett.}\ }\textbf {\bibinfo {volume} {97}},\ \bibinfo {pages}
  {076405} (\bibinfo {year} {2006})}\BibitemShut {NoStop}%
\bibitem [{\citenamefont {Yee}\ \emph {et~al.}(2010)\citenamefont {Yee},
  \citenamefont {Kotliar},\ and\ \citenamefont {Haule}}]{PuTe-Yee-2010}%
  \BibitemOpen
  \bibfield  {author} {\bibinfo {author} {\bibfnamefont {C.-H.}\ \bibnamefont
  {Yee}}, \bibinfo {author} {\bibfnamefont {G.}~\bibnamefont {Kotliar}}, \ and\
  \bibinfo {author} {\bibfnamefont {K.}~\bibnamefont {Haule}},\ }\href
  {\doibase 10.1103/PhysRevB.81.035105} {\bibfield  {journal} {\bibinfo
  {journal} {Phys. Rev. B}\ }\textbf {\bibinfo {volume} {81}},\ \bibinfo
  {pages} {035105} (\bibinfo {year} {2010})}\BibitemShut {NoStop}%
\bibitem [{\citenamefont {Gouder}\ \emph {et~al.}(2000)\citenamefont {Gouder},
  \citenamefont {Wastin}, \citenamefont {Rebizant},\ and\ \citenamefont
  {Havela}}]{PuSe-Gouder-2000}%
  \BibitemOpen
  \bibfield  {author} {\bibinfo {author} {\bibfnamefont {T.}~\bibnamefont
  {Gouder}}, \bibinfo {author} {\bibfnamefont {F.}~\bibnamefont {Wastin}},
  \bibinfo {author} {\bibfnamefont {J.}~\bibnamefont {Rebizant}}, \ and\
  \bibinfo {author} {\bibfnamefont {L.}~\bibnamefont {Havela}},\ }\href
  {\doibase 10.1103/PhysRevLett.84.3378} {\bibfield  {journal} {\bibinfo
  {journal} {Phys. Rev. Lett.}\ }\textbf {\bibinfo {volume} {84}},\ \bibinfo
  {pages} {3378} (\bibinfo {year} {2000})}\BibitemShut {NoStop}%
\bibitem [{\citenamefont {Durakiewicz}\ \emph {et~al.}(2004)\citenamefont
  {Durakiewicz}, \citenamefont {Joyce}, \citenamefont {Lander}, \citenamefont
  {Olson}, \citenamefont {Butterfield}, \citenamefont {Guziewicz},
  \citenamefont {Arko}, \citenamefont {Morales}, \citenamefont {Rebizant},
  \citenamefont {Mattenberger},\ and\ \citenamefont
  {Vogt}}]{PuTe-Durakiewicz-2004}%
  \BibitemOpen
  \bibfield  {author} {\bibinfo {author} {\bibfnamefont {T.}~\bibnamefont
  {Durakiewicz}}, \bibinfo {author} {\bibfnamefont {J.~J.}\ \bibnamefont
  {Joyce}}, \bibinfo {author} {\bibfnamefont {G.~H.}\ \bibnamefont {Lander}},
  \bibinfo {author} {\bibfnamefont {C.~G.}\ \bibnamefont {Olson}}, \bibinfo
  {author} {\bibfnamefont {M.~T.}\ \bibnamefont {Butterfield}}, \bibinfo
  {author} {\bibfnamefont {E.}~\bibnamefont {Guziewicz}}, \bibinfo {author}
  {\bibfnamefont {A.~J.}\ \bibnamefont {Arko}}, \bibinfo {author}
  {\bibfnamefont {L.}~\bibnamefont {Morales}}, \bibinfo {author} {\bibfnamefont
  {J.}~\bibnamefont {Rebizant}}, \bibinfo {author} {\bibfnamefont
  {K.}~\bibnamefont {Mattenberger}}, \ and\ \bibinfo {author} {\bibfnamefont
  {O.}~\bibnamefont {Vogt}},\ }\href {\doibase 10.1103/PhysRevB.70.205103}
  {\bibfield  {journal} {\bibinfo  {journal} {Phys. Rev. B}\ }\textbf {\bibinfo
  {volume} {70}},\ \bibinfo {pages} {205103} (\bibinfo {year}
  {2004})}\BibitemShut {NoStop}%
\bibitem [{\citenamefont {Wang}\ and\ \citenamefont
  {Yan}(2013)}]{TI-TopoHamiltonian-Wang}%
  \BibitemOpen
  \bibfield  {author} {\bibinfo {author} {\bibfnamefont {Z.}~\bibnamefont
  {Wang}}\ and\ \bibinfo {author} {\bibfnamefont {B.}~\bibnamefont {Yan}},\
  }\href {http://stacks.iop.org/0953-8984/25/i=15/a=155601} {\bibfield
  {journal} {\bibinfo  {journal} {Journal of Physics: Condensed Matter}\
  }\textbf {\bibinfo {volume} {25}},\ \bibinfo {pages} {155601} (\bibinfo
  {year} {2013})}\BibitemShut {NoStop}%
\bibitem [{\citenamefont {Mostofi}\ \emph {et~al.}(2008)\citenamefont
  {Mostofi}, \citenamefont {Yates}, \citenamefont {Lee}, \citenamefont {Souza},
  \citenamefont {Vanderbilt},\ and\ \citenamefont {Marzari}}]{Wannier90}%
  \BibitemOpen
  \bibfield  {author} {\bibinfo {author} {\bibfnamefont {A.~A.}\ \bibnamefont
  {Mostofi}}, \bibinfo {author} {\bibfnamefont {J.~R.}\ \bibnamefont {Yates}},
  \bibinfo {author} {\bibfnamefont {Y.-S.}\ \bibnamefont {Lee}}, \bibinfo
  {author} {\bibfnamefont {I.}~\bibnamefont {Souza}}, \bibinfo {author}
  {\bibfnamefont {D.}~\bibnamefont {Vanderbilt}}, \ and\ \bibinfo {author}
  {\bibfnamefont {N.}~\bibnamefont {Marzari}},\ }\href {\doibase
  10.1016/j.cpc.2007.11.016} {\bibfield  {journal} {\bibinfo  {journal}
  {Computer Physics Communications}\ }\textbf {\bibinfo {volume} {178}},\
  \bibinfo {pages} {685 } (\bibinfo {year} {2008})}\BibitemShut {NoStop}%
\bibitem [{\citenamefont {Kune\v{s}}\ \emph {et~al.}(2010)\citenamefont
  {Kune\v{s}}, \citenamefont {Arita}, \citenamefont {Wissgott}, \citenamefont
  {Toschi}, \citenamefont {Ikeda},\ and\ \citenamefont {Held}}]{W2W-Kunes}%
  \BibitemOpen
  \bibfield  {author} {\bibinfo {author} {\bibfnamefont {J.}~\bibnamefont
  {Kune\v{s}}}, \bibinfo {author} {\bibfnamefont {R.}~\bibnamefont {Arita}},
  \bibinfo {author} {\bibfnamefont {P.}~\bibnamefont {Wissgott}}, \bibinfo
  {author} {\bibfnamefont {A.}~\bibnamefont {Toschi}}, \bibinfo {author}
  {\bibfnamefont {H.}~\bibnamefont {Ikeda}}, \ and\ \bibinfo {author}
  {\bibfnamefont {K.}~\bibnamefont {Held}},\ }\href {\doibase
  10.1016/j.cpc.2010.08.005} {\bibfield  {journal} {\bibinfo  {journal}
  {Computer Physics Communications}\ }\textbf {\bibinfo {volume} {181}},\
  \bibinfo {pages} {1888 } (\bibinfo {year} {2010})}\BibitemShut {NoStop}%
\bibitem [{\citenamefont {Marzari}\ \emph {et~al.}(2012)\citenamefont
  {Marzari}, \citenamefont {Mostofi}, \citenamefont {Yates}, \citenamefont
  {Souza},\ and\ \citenamefont {Vanderbilt}}]{MLWF-RMP}%
  \BibitemOpen
  \bibfield  {author} {\bibinfo {author} {\bibfnamefont {N.}~\bibnamefont
  {Marzari}}, \bibinfo {author} {\bibfnamefont {A.~A.}\ \bibnamefont
  {Mostofi}}, \bibinfo {author} {\bibfnamefont {J.~R.}\ \bibnamefont {Yates}},
  \bibinfo {author} {\bibfnamefont {I.}~\bibnamefont {Souza}}, \ and\ \bibinfo
  {author} {\bibfnamefont {D.}~\bibnamefont {Vanderbilt}},\ }\href {\doibase
  10.1103/RevModPhys.84.1419} {\bibfield  {journal} {\bibinfo  {journal} {Rev.
  Mod. Phys.}\ }\textbf {\bibinfo {volume} {84}},\ \bibinfo {pages} {1419}
  (\bibinfo {year} {2012})}\BibitemShut {NoStop}%
\end{thebibliography}
\end{document}